\documentclass[superscriptaddress,aps,preprint,longbibliography]{revtex4-1}
\usepackage{graphicx} 
\usepackage{amsmath}
\usepackage[utf8]{inputenc}
\usepackage{booktabs}
\usepackage{enumitem}
\graphicspath{{./pictures/}}
\begin{document}
\title{Melancholia States of the Atlantic Meridional Overturning Circulation}
\author{Johannes Lohmann}
\email{johannes.lohmann@nbi.ku.dk}
\affiliation{Physics of Ice, Climate and Earth, Niels Bohr Institute, University of Copenhagen, Denmark}
\author{Valerio Lucarini}
\affiliation{School of Computing and Mathematical Sciences, University of Leicester, Leicester, UK}

\begin{abstract}
The Atlantic Meridional Overturning Circulation (AMOC) is a much studied component of the climate system, because its suspected multistability is associated with tipping behaviour yielding potentially large regional and global climatic impacts. In this paper we investigate the global stability properties of the system using an ocean general circulation model. We construct an unstable AMOC state, i.e., an unstable solution of the flow that resides between the stable regimes of a vigorous and collapsed AMOC. Such a solution, also known as a Melancholia or edge state, is a dynamical saddle embedded in the boundary separating the competing basins of attraction. It is physically relevant since it lies on the most probable path of a noise-induced transition between the two stable regimes, and because tipping occurs when one of the attractors and the Melancholia state collide. Its properties may thus give hints towards physical mechanisms and predictability of the critical transition. We find that while the AMOC Melancholia state as viewed from its upper ocean properties lies between the vigorous and collapsed regimes, it is characterized by an Atlantic deep ocean that is fresher and colder compared to both stable regimes. The Melancholia state has higher dynamic enthalpy than either stable state, representing a state of higher potential energy that is in good agreement with the dynamical landscape view on metastability. 
\end{abstract}

\maketitle

\section{Introduction}
The Earth's climate history features periods of smooth changes interlaced with rapid variations likely due to critical transitions \citep{Westerhold2020,Ghil2020,Boers2022,Rousseau2023}, the latter being sometimes associated with mass extinctions events and fundamental changes in the biosphere \citep{Lenton2013,Alsulami2024,Algeo2024}. Critical transitions in the Earth system arise when the states of different large-scale climatic sub-components lose stability. There is widespread concern that ongoing climate change could indeed trigger the onset of transitions in various so-called tipping elements of the climate system. These include the Amazon Forest, the Atlantic Meridional Overturning Circulation (AMOC), the Greenland ice sheet, and the Permafrost, among others \citep{LEN08,ArmstrongMcKay2022}.  Hence it is nowadays rather common to refer to the ongoing impact of anthropogenic activities on the climate system as the \textit{climate crisis} \citep{Lucarini2023}.

The Earth system consists of a variety of sub-systems that evolve on time scales spanning many orders of magnitude \citep{Ghil2020,VDH21}. The sub-systems are coupled by various physical processes, and as a result the combined Earth system is a multi-scale dynamical system. But it has long been argued, along the lines of the Hasselmann program \citep{Imkeller2001}, that in some circumstances these time scales may be treated as sufficiently separated, so that a sub-system evolving on a particular time scale can be considered to be forced by stochastic variability in sub-systems that evolve on much faster time scales \citep{HAS76}. This motivates why stochastic parameterizations are needed to improve the performance of weather forecast and climate models \citep{Berner2017}. On the other extreme, if there are sub-systems that evolve on even much longer time scales than the element in question, the effect of these sub-systems may be considered a slowly-varying, quasi-adiabatically changing control parameter on the time scale of interest \citep{Saltzman2001}. If one relaxes the assumption of time-scale separation, the Hasselmann program requires a careful re-examination, because rates of change become non-negligible, the stochastic forcing is not white, and the dynamics includes non-markovian contributions \citep{Lucarini2023}.

As clarified in \cite{ASH12}, tipping behaviour can result, by and large, from three different mathematical mechanisms, which are often co-existent in real-life systems: 1) classical {\it bifurcations} associated with quasi-adiabatic modulations of the parameters of the system (B-tipping); 2) transitions occurring specifically as a result of non-negligible {\it rates} of change of one or more system parameters (R-tipping); and 3) {\it stochastic} perturbations inducing transitions between competing modes of operation of the system (N-tipping).

These tipping mechanisms all rely on an underlying multistability of the system, which originates in nonlinear dynamical laws associated with positive feedbacks, leading to the existence of multiple competing stable states. A notable example of multistability in the Earth system is the AMOC, the stability properties of which have long been studied beginning with the landmark contribution by Stommel \citep{Stommel1961}, who highlighted using a simple box model that the competing (destabilizing) salinity and (stabilizing) temperature advection feedbacks can result in more than one competing state. Here, the freshwater flux into the ocean is the natural parameter controlling whether the system is in a monostable or bistable configuration. One state - resembling the present-day AMOC conditions - is characterized by a vigorous interhemispheric circulation ('ON' state). The competing state instead features weak or absent large scale flow ('OFF' state). The driving mechanisms, controls, and stability of the AMOC are being actively studied \citep{Kuhlbrodt2007,Weijer2018}, especially because variations in the intensity and features of the AMOC can have key impact on the regional and even global climate \citep{Rahmstorf2002,Liu2020}, and are thought to have played a major role in large climatic variations in the Pleistocene \citep{Broecker1990,Barker2021}. 

A wealth of studies performed using models of different complexity support the overall picture proposed by Stommel \citep{Rahmstorf2005,Hawkins2011}, and observations of so-called early-warning signals associated with B-tipping indicate that the ongoing climate change might be leading to a loss of stability of the current active state of the AMOC, i.e., a loss of the present intense interhemispheric ocean heat transport \citep{Boers2021,Ditlevsen2023,vanWesten2024}. 
But it is worth noting that the real-world Atlantic comprises different interconnected turbulent circulation systems \citep{ROQ22}, which makes it hard to actually measure the AMOC and its component that is most affected by climate change and positive feedbacks. As a result, there is considerable uncertainty in how large the risk of an abrupt AMOC collapse is, what the preceding warning signs would be, and what the potential impacts are. As an example showcasing the complexity of the AMOC beyond the Stommel '61 picture, a recent modelling study \citep{Lohmann2024} indicates that the AMOC might possess more than just two competing states for certain values of the system parameters: several distinct variants are found within both regimes of strong and weak circulation, associated with a rather complex bifurcation diagram and a web of possible transitions between the various competing states as parameters are varied. The presence of such a complex dynamical landscape suggests that the above-mentioned early-warning signals do not necessarily indicate that a loss of stability of the current state will lead to the AMOC falling into the 'OFF' state. While early-warning signals investigate local properties of the system, it is clear that a more global view on the stability landscape of the system is helpful.

\subsection{Multistability and Edge States}
In multistable systems, the asymptotic state depends on the initial conditions, with different subsets of initial conditions (basins of attraction) evolving towards different stable states (attractors). If the system is initialised on the boundary between neighbouring basins of attraction, the state evolves towards a special asymptotic set, the edge state, which is a saddle that separates the competing attractors. Small perturbations away from the saddle result in the system converging to either attractor with probability one. 

Edge states in physical systems were first studied in the context of classical fluid dynamical configurations like the plane Couette flow and the pipe flow \citep{Schneider2008,Schneider2009,deLozar2012}, even if, rigorously speaking, such systems possess just one true attractor (the laminar flow), which is co-existing, in the high-forcing regime, with an extremely long-lived turbulent transient \citep{Hof2006,Hof2008}. The so-called edge tracking algorithm \citep{Skufca2006,Vollmer2009} has been a key tool to construct edge states, which are otherwise inaccessible using standard forward numerical integration because of its unstable nature. 

The edge tracking algorithm has also allowed the computation of edge states in nontrivial climate models featuring highly turbulent behaviour and co-existing warm and snowball states \citep{Lucarini2017}. There, the dynamical saddle was found to be a chaotic state, and in this particular case was referred to as Melancholia (M) state, a nomenclature we will follow hereafter as the dynamics in the system we consider here is also chaotic. The presence of a chaotic saddle was also shown to be linked to the presence of a fractal basin boundary between the competing states, featuring a co-dimension strictly smaller than one \citep{Vollmer2009,Bodai2020}. As a result, one finds a vast region of phase space where the so-called predictability of the second kind \citep{lorenz1975climatic} is extremely limited: applying arbitrarily small perturbations to the initial conditions can lead to system ending up in a different asymptotic state \citep{Mehling2024}.  

The M state is physically relevant because tipping behaviour associated with quasi-adiabatic parametric modulations occur when one of the attractors and the M state collide. Hence, its properties may give hints towards physical mechanisms and predictability of the critical transition. The M state may also be key to interpret some cases of rate-induced tipping, where a temporary attraction to the M state can occur \citep{WIE11}, resulting in long transients \citep{LOH21b,LOH21}. Finally, the M state can be of great relevance in the case of noise-induced tipping. The presence of stochastic forcing allows for transitions between the different basin of attraction, leading to the phenomenon of metastability. Graham's field theory \citep{Graham1987,Graham1991} indicates that in the weak-noise limit transitions between competing basins of attraction can be described - both in terms of time statistics and paths - by large deviation theory \citep{Touchette2009}. Transition paths are concentrated around a special trajectory known as instanton, which connects the attractors of the system with the M state living on the basin boundary. The M state can thus be considered the gateway of noise-induced transitions. Identifying the physical characteristics of an M state or instanton can be useful as an indicator that a transition is on the way, and it can help identify the physical mechanisms and perturbations that promote a transition.

\subsection{This paper}
Evidence for regime transitions in geophysical flows that may be understood as stochastic transitions has been found for oceanic western boundary currents \citep{SCH01}, zonal-blocking transitions in the atmosphere \citep{PLA93}, and geomagnetic reversals \citep{GLA95,VAL03}. Previous work has identified instantons and edge states from simulations of geophysical flows, such as for atmospheric jet dynamics \citep{BOU19} and the snowball Earth transition \citep{Lucarini2019,Lucarini2020,Margazoglou2021}. But the M state associated with regime transitions of the AMOC has only been started to be investigated recently \citep{boerner2024,Mehling2024}. A case for time scale separation, stochastic forcing and spontaneous transitions of the AMOC has been made previously \citep{CES94, DRI13, KLE15, CAS20}. 
Recently, the possibility of a spontaneous AMOC collapse has been observed using a rare-event algorithm \citep{Giardina2006,Ragone2018}, where a coupled climate model was used, which, unlike a truly stochastic atmosphere, explicitly simulates the turbulent flow of the atmosphere \citep{CIN24}. 


Here we analyze the pathway to a hypothetical collapse of the AMOC by computing the M state that lies between the vigorous and collapsed AMOC regimes, based on the recently constructed stability landscape of a global ocean model under North Atlantic (NA) freshwater forcing \citep{Lohmann2024}. Its physical characteristics are analyzed in order to assess whether they can be harnessed to detect if a spontaneous collapse of the AMOC is underway, and to elucidate the physical mechanisms that would facilitate such a transition. 

The structure of the paper is as follows. First, we present a qualitative survey of the co-existing attractors and their basins of attraction based on a large ensemble simulation  (Sec.~\ref{sec:basins}). Next, simulations are performed that precisely locate the basin boundary separating the vigorous and collapsed AMOC regimes, and, from a given position on this basin boundary, edge tracking is performed (Sec.~\ref{sec:edge_tracking}). The physical properties of the computed M state are analyzed in Sec.~\ref{sec:edge_state}, and it is discussed in Sec.~\ref{sec:fingerprint} how this can be used to detect a fingerprint of an AMOC collapse. 
Conclusions are drawn in Sec.~\ref{sec:conclusions}. Finally, Appendix~\ref{Appe} contains details on the numerical model used.

\section{Results}

\begin{figure}
\includegraphics[width=0.99\textwidth]{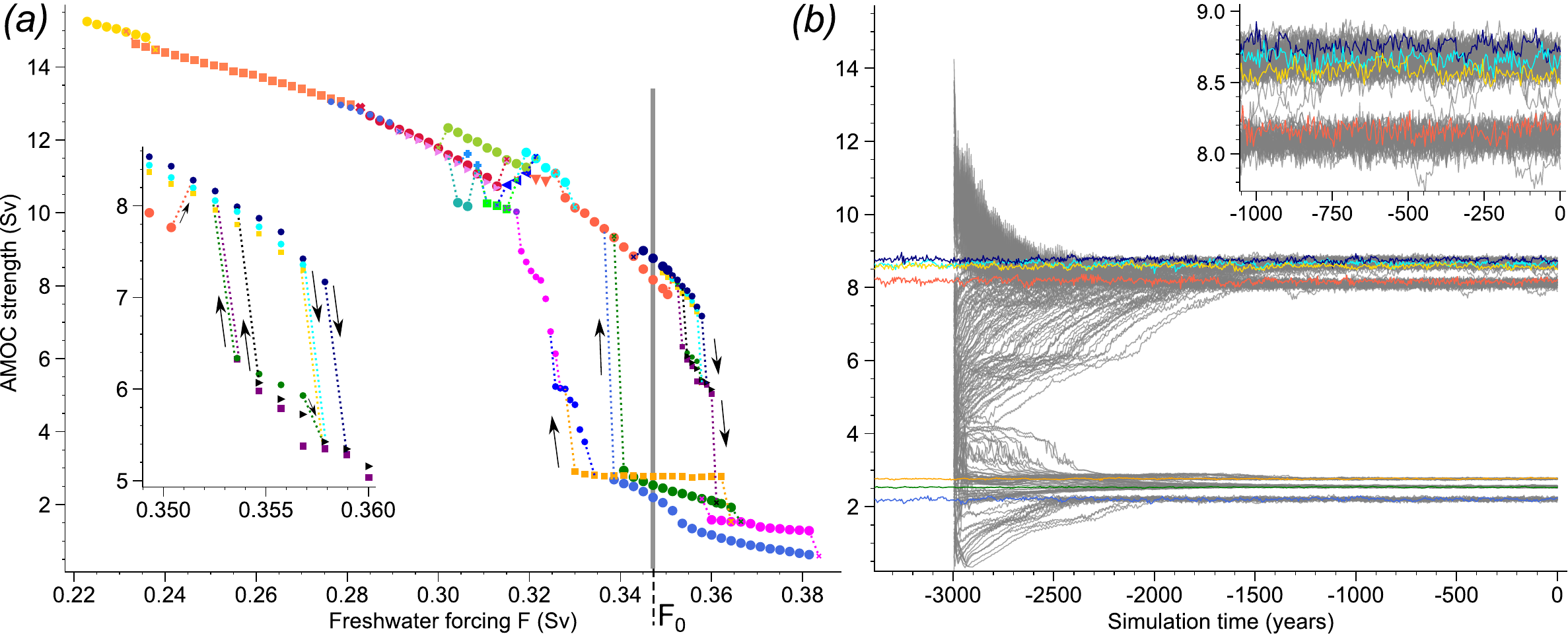}
\caption{\label{fig:sections} 
{\it (a)} Stability diagram of AMOC in the ocean model, as obtained in \citet{Lohmann2024}. Symbols of different color indicate the different branches of attractors. Dotted lines and their color indicate which attractor branch collapses onto which other branch. The vertical line indicates the parameter value at which we performed ensemble simulations that are presented in this study. The inset shows a zoom-in on the same stability diagram, highlighting the smaller-scale hysteresis loop with very close-by co-existing vigorous and partially-collapsed states. 
{\it (b)} Maximum AMOC strength time series of the individual realizations of the ensemble (gray) at fixed forcing $F_0$ with $N=199$ members, along with longer steady-state simulations (at least 10~000 years) started close to each of the co-existing attractors (colored time series). The inset details the subset of trajectories that converged onto the two bands of vigorous AMOC attractors. 
}
\end{figure}

\subsection{Survey of the basin structure}
\label{sec:basins}

We perform simulations with the primitive equation finite-difference global ocean model {\it Veros} \citep{HAE18,HAE21} in a coarse-resolution setup (see Appendix~\ref{Appe} for a description of the model) to enable the long simulation times required to compute an unstable M state. 
The AMOC in the model has been shown to display a high degree of multistability \citep{Lohmann2024}, as seen in the bifurcation diagram in Fig.~\ref{fig:sections}a, where symbols of same shape and color indicate the different branches of attractors and the control parameter on the horizontal axis is the NA freshwater forcing $F$. 
At the TP corresponding to the critical value $F\approx0.36$~Sv the stable states with a vigorous AMOC disappear and the system undergoes a transition to a collapsed AMOC state. 
This critical value is comparable to many other modeling studies \citep{RAH05,vanWesten2024}, but much higher than current \citep{BEV19} and projected \citep{ASC19} NA freshwater forcing, assuming it derives from Greenland ice melt alone. 
The actual AMOC TP may, however, be lower due to model biases \citep{LIU17}, due to positive future anomalies in precipitation minus evaporation over the ocean basin \citep{BYR15,ELB22}, and due to the compounding effect of the NA thermal forcing as a result of global warming. 
Further, unlike in hysteresis experiments with state-of-the-art coupled models \citep{VWE23}, the range of multistability in our model is very narrow. We believe this is specific to the particular configuration of this model, where salinity relaxation boundary conditions are used that yield a relatively weak, destabilizing salt-advection feedback. Such a feedback is positive, so having it in a weaker form reduces the instability of the model. Note that for such a narrow range of multistability, there is only a limited ``irreversibility'' of the TP. A reversal of forcing after crossing the TP, for instance due to depletion of the Greenland ice sheet within a millennium at such high levels of meltwater discharge \citep{ASC19}, would lead to a relatively quick recovery of the AMOC.  Due to the limitations of our model, we do not aim to make assessments about the existence and reversibility of the AMOC TP in the real world. Instead, the model is well suited for a theoretical study of the unstable state between the AMOC ON and OFF regimes.

\begin{figure}
\includegraphics[width=0.85\textwidth]{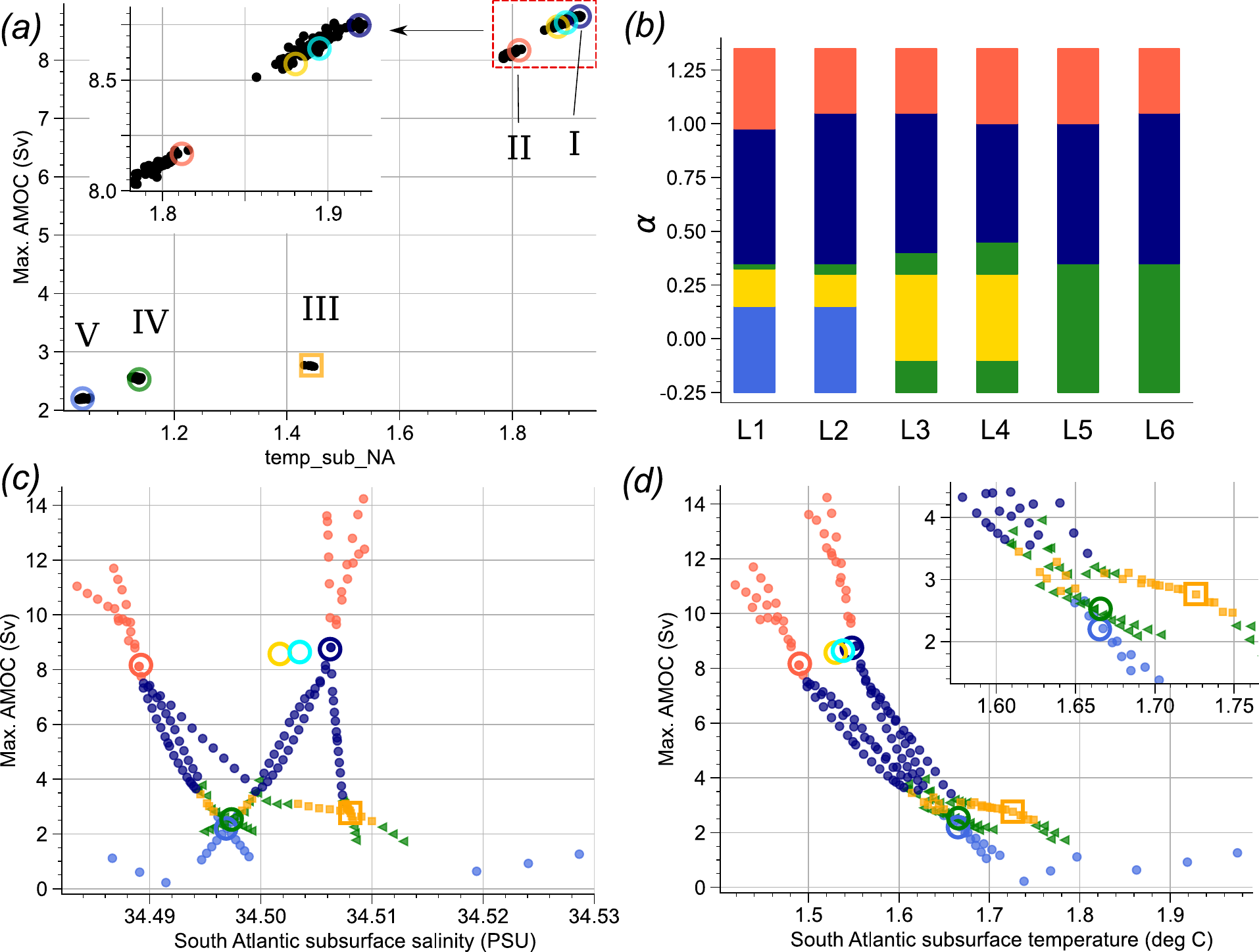}
\caption{\label{fig:basin_structure} 
{\it (a)} Two-dimensional projection of the asymptotic states of the ensemble members onto the maximum AMOC strength and the NA sub-surface temperature, averaged over the last 400 years of the simulations (black dots). The open, colored circles are 2000-year mean values on the co-existing attractors. We labeled the five of the attractors (I to V) that have been used to construct the ensemble simulations. The inset is a closeup of the bands that comprise attractors I and II. 
{\it (b)} Sequence of basins of attraction along six different lines L1-L6 across phase space, defined by interpolating between pairs of states on the attractors II-V, I-V, I-III, II-III, II-IV, and I-IV, respectively. 
{\it (c,d)} Initial states of the ensemble members projected onto two scalar observables, and color coded according to which attractor the simulations converge to (for color coding see Fig.~\ref{fig:sections}). Projection onto maximum AMOC strength and South Atlantic subsurface salinity {\it (c)} and temperature {\it (d)}. 
The inset in {\it (d)} shows how the basins overlap in this projection close to the location of the AMOC OFF attractors.
}
\end{figure}

To this end, we focus on the forcing parameter $F=F_0=0.347$~Sv, which lies in the middle of the main hysteresis loop (vertical line in Fig.~\ref{fig:sections}a). At $F_0$ there is a cluster of three distinct states featuring a very weak AMOC, as well as a cluster of states featuring a vigorous AMOC. This second cluster is further divided into two bands of states with an average strength of approximately 8.2 and 8.7 Sv, respectively (see colored time series in Fig.~\ref{fig:sections}b and inset). For the band with slightly larger AMOC strength, three distinct attractors have been identified in \citet{Lohmann2024} (see inset Fig.~\ref{fig:sections}a). In the lower band only one attractor was identified, but both the upper and lower bands might contain more attractors. For slightly higher $F$, an additional cluster of three partially-collapsed states forms another smaller-scale hysteresis loop (inset in Fig.~\ref{fig:sections}a), in agreement with the hierarchical structure of the dynamical landscape of the climate system conjectured in \citep{Margazoglou2021}. In terms of different macroscopic observables, states within the clusters may be separated more clearly. The open circles in Fig.~\ref{fig:basin_structure}a show a two-dimensional projection of the mean states on the attractors, where the cluster of collapsed AMOC states is seen to split up in terms of the NA subsurface temperature. Similarly, the two bands of vigorous states are well-separated in terms of other observables (see e.g. open symbols in Fig.~\ref{fig:basin_structure}c).

As shown in \citet{Lucarini2017} and \citet{Mehling2024}, computing the M state of a bistable system is relatively straightforward once one has a working edge tracking algorithm in place, the main challenge being the need to initialize the procedure with two very nearby initial conditions on the two opposite sides of the basin boundary. Against intuition, the procedure works even in the case of M states, where the basin boundary has codimension strictly smaller than one - and possibly very close to zero. In the case studied here, the situation is considerably more complicated because the system has at least seven distinct chaotic attractors, which leaves the door open for a large variety of geometrical configurations for the basins of attraction and of the features of the basin boundaries. 

Hence, as preliminary step to computing an M state, we perform a survey of the basins of attraction and their boundaries for the co-existing attractors, using ensemble simulations with a large variety of initial conditions. This is achieved by choosing states along several straight lines in phase space, yielding a handful of one-dimensional 'sections' of phase space. The lines we construct are defined by two states $A$ and $B$ on a pair of attractors, where one attractor features a vigorous AMOC and the other a collapsed AMOC. The states are chosen as the model snapshot at the last time step of the long steady-state simulations that were performed at fixed forcing values in \citet{Lohmann2024}, and thus approximately correspond to one individual point in each of the distinct attractors. We then perform a linear interpolation of the entire set of state variables and construct 32 equally spaced initial conditions $C$: 
\begin{equation}
\label{eq:lines}
C = \alpha A + (1 - \alpha) B, \,\,\,\, \text{where } \alpha = (i-5)/20, \,\,\,\, \text{and } i = \{0,1,..,31\}.
\end{equation}
The values $i=\{5,6,..,25\}$ correspond to an interpolation between the pair of states, and considering larger and smaller values of $i$ allows us to sample a larger variety of initial ocean states. 

To construct initial conditions according to Eq.~\ref{eq:lines}, we choose the three weak AMOC attractors and two of the four vigorous AMOC attractors (dark blue and red time series in Fig.~\ref{fig:sections}b). We are unable to consider all four vigorous attractors (cyan and light yellow are excluded) because of computing limitations. Nonetheless, we are able to cover both bands of vigorous attractors, see above. This yields six pairings of a vigorous and a collapsed attractor and an ensemble comprising initial conditions lying on six different lines L1-L6 in phase space (see Fig.~\ref{fig:basin_structure}a,b for definition of the attractors and lines). Figure~\ref{fig:sections}b shows how the ensemble trajectories (gray) converge to the different attractors (colored time series). Within the simulation time of 3~000 years it can be determined from the AMOC strength which attractors each ensemble member converges to, with the exception of not being able to clearly distinguish the close-by states in the uppermost band of the vigorous AMOC attractors (inset Fig.~\ref{fig:sections}b). 

Figures~\ref{fig:basin_structure}a show a projection of the asymptotic states of the ensemble members (black circles) defined by the mean AMOC strength and NA sub-surface temperature. The ensemble converges to the two distinct bands of vigorous AMOC attractors, but within the bands a distinction of attractors is difficult (inset Fig.~\ref{fig:basin_structure}a). Longer simulation times would be required to converge even closer to the attractors and to allow for longer time averages that yield more precise mean values by eliminating the centennial-scale variability. As a result, we neglect some of the fine structure and in the following treat the uppermost band as one attractor color coded in dark blue. 

Pairing the initial conditions of the ensemble members with their asymptotic states gives useful information on the basin boundaries for $F=F_0$. Figure~\ref{fig:basin_structure}c,d shows two-dimensional projections of the initial conditions that are color coded according to which attractor they converge to. The sequence of basins along the individual lines in phase space that constitute the ensemble is shown as a function of $\alpha$ (Eq.~\ref{eq:lines}) in Fig.~\ref{fig:basin_structure}b. The connection from vigorous to collapsed AMOC regimes is always via the sequence dark blue basin - green basin. These are the basins of the two attractors we refer to as 'ON' and 'OFF' hereafter. Although based on a limited number of initial conditions and very sparse exploration of an extremely high-dimensional phase space, this may indicate that as a result of how the basins are geometrically arranged in phase space, there is a specific path that needs to be traversed in a transition between the vigorous and collapsed AMOC regimes, regardless of which specific vigorous and collapsed attractors are the starting and end points. 

We remark that the dark blue and green states are not those one would naively choose by looking at the AMOC intensity as indicator of proximity: the dark blue attractor has a more vigorous circulation than the red one; and the AMOC of the green attractor is slightly weaker than that of the yellow attractor. This further indicates that the AMOC intensity alone does not necessarily provide sufficient information to understand the detailed dynamics of such a high-dimensional flow. 

\subsection{Basin Boundaries and Edge Tracking}
\label{sec:edge_tracking}

To precisely locate the basin boundary, we follow the initial step of the edge tracking algorithm and begin with an arbitrary pair of initial conditions that converge to ON and OFF, respectively. We bisect the two initial conditions and use the resulting state as initial condition for a new simulation, where it is determined whether the trajectory converges to ON or OFF. Next, we bisect the newly constructed initial state and the previous initial condition that converged to the opposite attractor, and use the resulting state as initial condition for the next simulation. This procedure is repeated for $M$ iterations, which yields two very close-by initial conditions on opposite sides of the basin boundary. 

\begin{figure}
\includegraphics[width=0.99\textwidth]{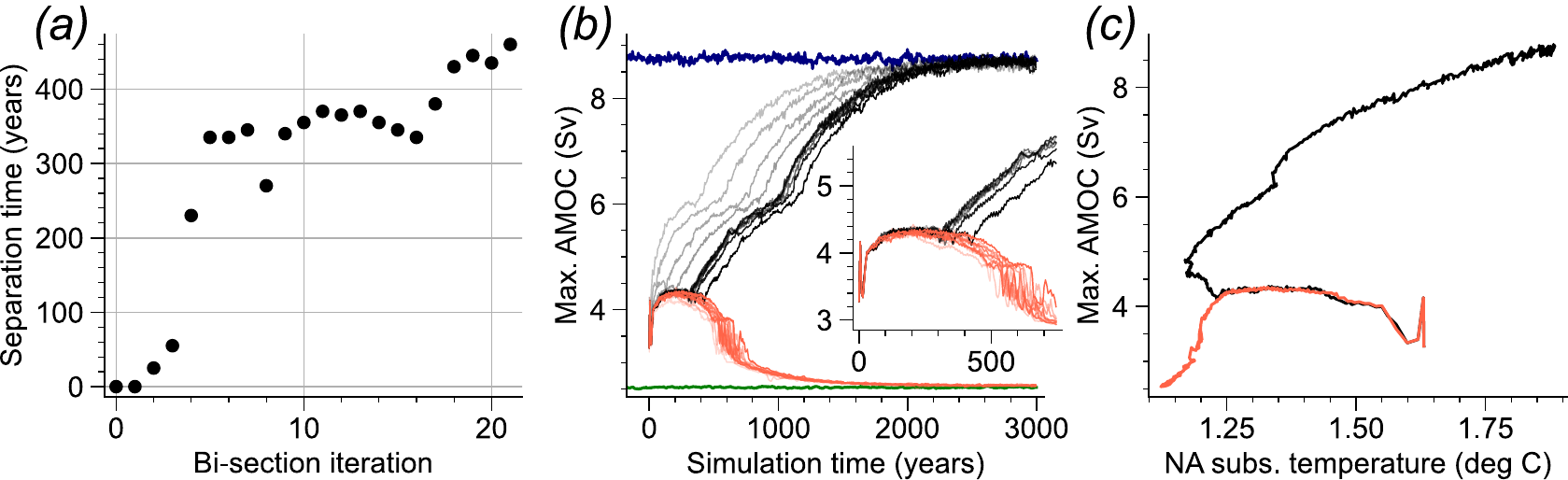}
\caption{\label{fig:boundary_bisection} 
Repeated bisections of initial conditions are performed, such that a pair of very close-by initial conditions is obtained where one reaches the ON attractor, and the other the OFF attractor.  
{\it (a)} For each iteration of the bisection procedure, we estimate the time until the pair of trajectories has separated significantly. This is defined by the first time the trajectories deviate in AMOC strength by more than 0.2 Sv, as estimated from the time series smoothed by a 20-year Gaussian kernel filter. 
{\it (b)} Time series of the AMOC strength of the individual realizations during the bi-section procedure, drawn with increasing contrast for increasing iteration number. The inset is a close-up on the time series after the fifth iteration. 
{\it (c)} Two-dimensional projection onto the AMOC strength and NA sub-surface temperature of the pair of trajectories at the final iteration. 
}
\end{figure}

Figure~\ref{fig:boundary_bisection} shows one instance of this procedure starting from a pair of initial conditions on the line L3 that belong to the dark blue and green basins, respectively, as defined in Fig.~\ref{fig:basin_structure}. At each iteration, we measure the time elapsed until the pair of trajectories separates appreciably in terms of AMOC strength (Fig.~\ref{fig:boundary_bisection}a), after which they converge to opposite attractors. This time increases rapidly for the first few iterations and then stabilizes at several hundred years (Fig.~\ref{fig:boundary_bisection}b). The final pair of trajectories is very close-by at $\alpha = \frac{6453063}{16777216} = 0.384632528$ and $\alpha = \frac{806633}{2097152} = 0.384632587$, where the large integers are due to the $M=21$ bisections from the initial states at $\alpha=\frac{1}{2}$ and $\alpha=\frac{1}{3}$. The trajectories evolve closely in phase space for almost 500 years before separating (Fig.~\ref{fig:boundary_bisection}c), indicating they are closely following the stable manifold of the M state (the basin boundary) before the transversal instability takes over \citep{Lucarini2017}. These trajectories are then used to initialize the main component of the edge tracking procedure.

We repeatedly apply the bisection method above to new initial states that are created by a relatively short time evolution of the two closest initial conditions on opposite sides of the boundary. At each iteration $i$, pairs of initial conditions that lie close-by on opposite sides of the basin boundary are constructed by a fixed number of $k$ bisections, in the following way. Starting from a pair of initial states $\mathcal{A}$ and $\mathcal{B}$ that converge to the ON and OFF attractor, respectively, a bisection is performed to obtain a new initial state $\mathcal{C} = (\mathcal{A} + \mathcal{B})/2$. $\mathcal{C}$ is then evolved from time $t_0$ to $t_2$, chosen such that we can decide with certainty from a threshold on the AMOC strength whether the trajectory will converge to ON or OFF. If $\mathcal{C}$ converges to ON we assign $\mathcal{C} \rightarrow \mathcal{A}$, and if it converges to OFF we set $\mathcal{C} \rightarrow \mathcal{B}$. Thereafter a new initial condition $\mathcal{C} = (\mathcal{A} + \mathcal{B})/2$ is constructed, and the procedure is repeated until $k$ bisections are reached. The trajectories associated with the states $\mathcal{A}$ and $\mathcal{B}$ are saved and a new iteration $i+1$ of the edge tracking procedure is spawned by using the model snapshots obtained after the time evolution of $\mathcal{A}$ and $\mathcal{B}$ until $t_1 < t_2$ as starting initial states for a new round of $k$ bisections. $t_1$ is chosen such that the trajectories begin to separate, but are not too far from the boundary yet. 

\begin{figure}
\includegraphics[width=0.85\textwidth]{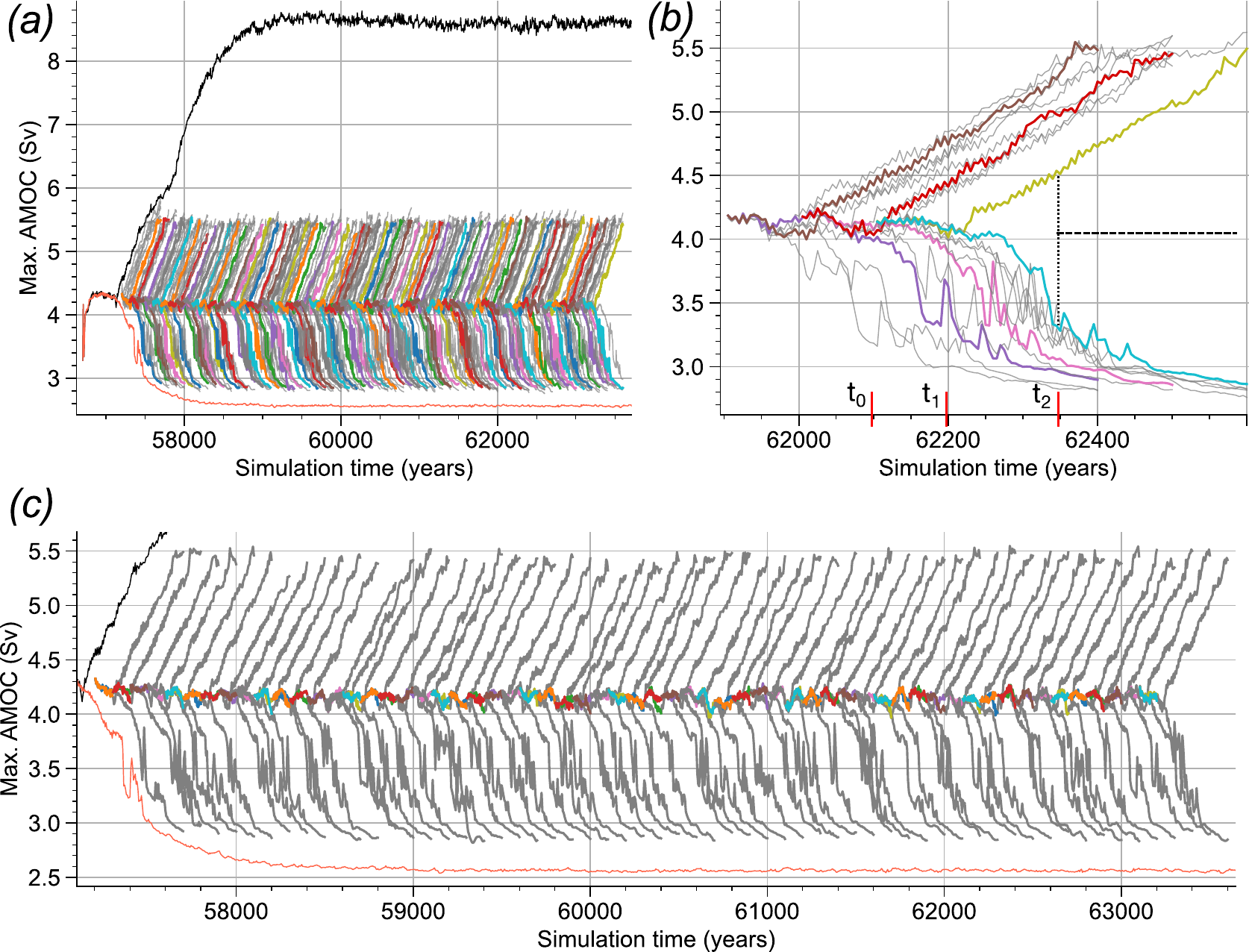}
\caption{\label{fig:timeseries_all} 
{\it (a)} All simulations performed during one instance of the edge tracking procedure with $N=60$ iterations, shown as time series of the AMOC strength. The longer red and black trajectories are the simulations from which the initial state is chosen. Thereafter, the edge tracking algorithm produces subsequent generations of trajectories, where for each iteration the colored time series correspond to the pair of simulations with the closest-by initial conditions that converge to opposite attractors. The gray trajectories are all other simulations. 
{\it (b)} Closeup of the time series for three subsequent iterations. The threshold used to decide whether a simulation is converging to ON or OFF is shown with the black, dashed line. 
{\it (c)} As in {\it (a)}, but only the final pairs of each iteration are shown, and only the portion of the time series until $t_1$ (when a new iteration is branched off) is given in color.
}
\end{figure}

If the bisections yield initial states close enough to the boundary, at each iteration the model states move a little closer towards the M state along the basin boundary, as we are shadowing a true trajectory that is moving along the attractive, stable manifold of the M state. Figure~\ref{fig:timeseries_all} shows results of the edge tracking procedure initiated from model snapshots taken shortly after the separation of the trajectories shown in Fig.~\ref{fig:boundary_bisection}c. We choose $k=7$ bisections, since at the given forcing $F_0$ this is sufficient to generate pairs of trajectories that evolve along the boundary for a longer period of time (Fig.~\ref{fig:boundary_bisection}a). Further, $t_1 = 100$~years and $t_2 = 250$~years are used, and all trajectories are continued until $t_3=500$~years to check whether they indeed continue on the path to the attractor that was determined at $t_2$. We use $t_2 < t_3$ for computational reasons, such that two realizations can be simulated in parallel. All time series generated during the edge tracking are shown in Fig.~\ref{fig:timeseries_all}a. The algorithm produces pairs of trajectories that reliably evolve close to each other for roughly a century (Fig.~\ref{fig:timeseries_all}b,c), after which they quickly separate, allowing us to detect reliably at $t_2$ whether they converge to ON or OFF. Here we used trial and error to roughly match $t_1$ to the separation time of the closest pairs of trajectories. 
 
\begin{figure}
\includegraphics[width=0.8\textwidth]{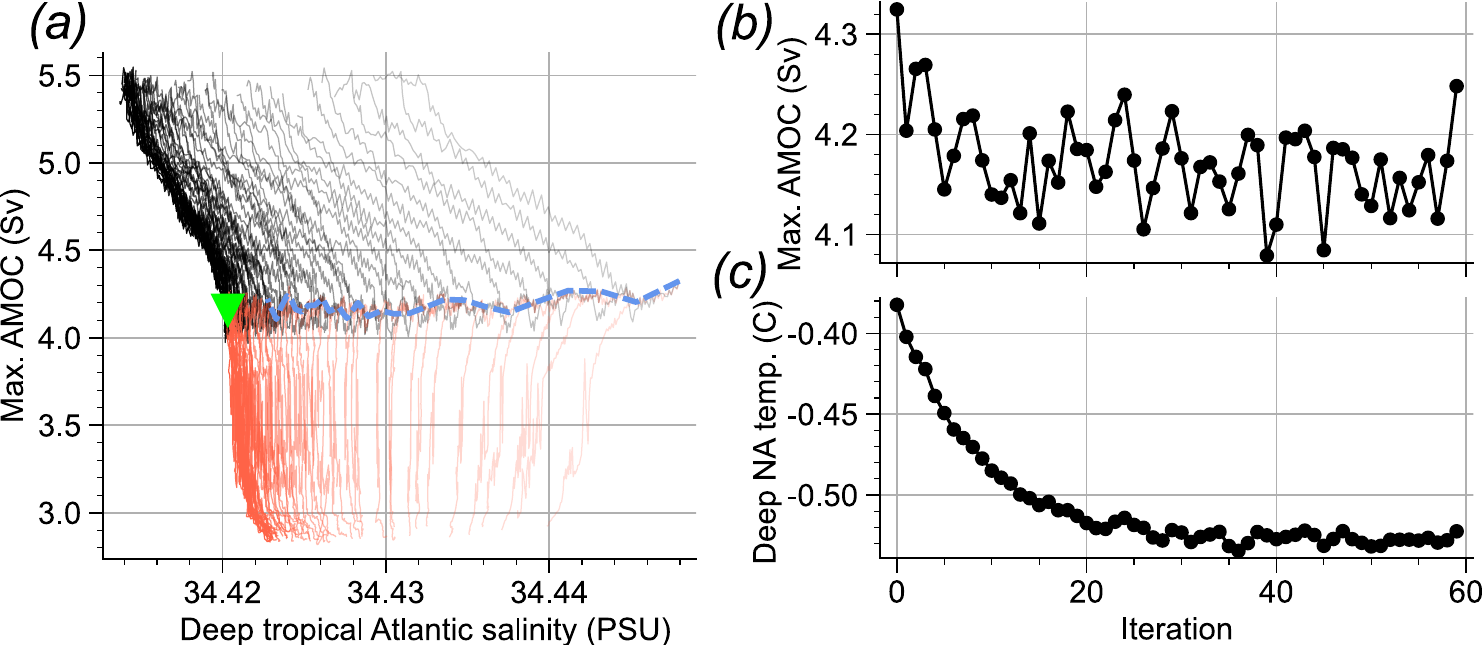}
\caption{\label{fig:convergence} 
{\it (a)} Pairs of trajectories in 2D projection of AMOC strength and tropical Atlantic deep ocean salinity for all edge tracking iterations. The iterations are progressing from right to left, which is further indicated by increase in contrast of the trajectories. The red (black/gray) trajectories are the simulations that converge to OFF (ON). The average initial states of the trajectory pairs converge to the estimated M state, shown by the green triangle. The estimated basin boundary is given by the blue dashed line.
{\it (b)} Average state in terms of AMOC strength of the final pair of trajectories for each iteration, obtained from the average model state of the first 5 simulation years, averaged over the trajectory pairs. 
{\it (c)} Same, but for the NA deep ocean temperature.
}
\end{figure}

It is apparent from Fig.~\ref{fig:timeseries_all} that the AMOC strength does not provide any useful indication of the motion along the basin boundary, as by looking at its time series we are unable to elicit whether the true trajectory we are shadowing has reached the M state.  Figure~\ref{fig:convergence}a shows a two-dimensional projection of the trajectories of the pairs of closest initial conditions for all iterations. The evolution of the AMOC strength is very similar for all iterations, but other observables related to the deep ocean water properties show a clear trend as the edge tracking procedure progresses. During early iterations, subsequent trajectories in Fig.~\ref{fig:convergence}a are spaced relatively far apart along the horizontal axis, as they move quickly towards initial conditions with lower Atlantic deep ocean salinity. For later iterations subsequent trajectories get closer and closer together, until they converge to a stationary region in the two-dimensional projection.

By averaging over the first 5 simulation years of a pair of trajectories at a given iteration, we estimate the shadowed state on the basin boundary. As the edge tracking iteration proceeds, such a state moves along the boundary (blue dashed line in Fig.~\ref{fig:convergence}a) and asymptotically tends towards the M state (green triangle, estimated from the average state of the last 10 iterations), showing that the M state is indeed a relative attractor for the basin boundary \citep{Lucarini2017}. The AMOC strength of the shadowed state does not change much after a few iterations (Fig.~\ref{fig:convergence}b). But there is an approximately exponential relaxation of the slow variables associated with the deep ocean, converging to a steady state after 30-40 iterations (Fig.~\ref{fig:convergence}c). Given $t_1 = 100$~years, the resulting relaxation time is in good agreement with the typical multi-millennial time scale until the model equilibrates onto one of its attractors \citep{Lohmann2024}.

\subsection{Characteristics of the Edge State}
\label{sec:edge_state}

\begin{figure}
\includegraphics[width=0.85\textwidth]{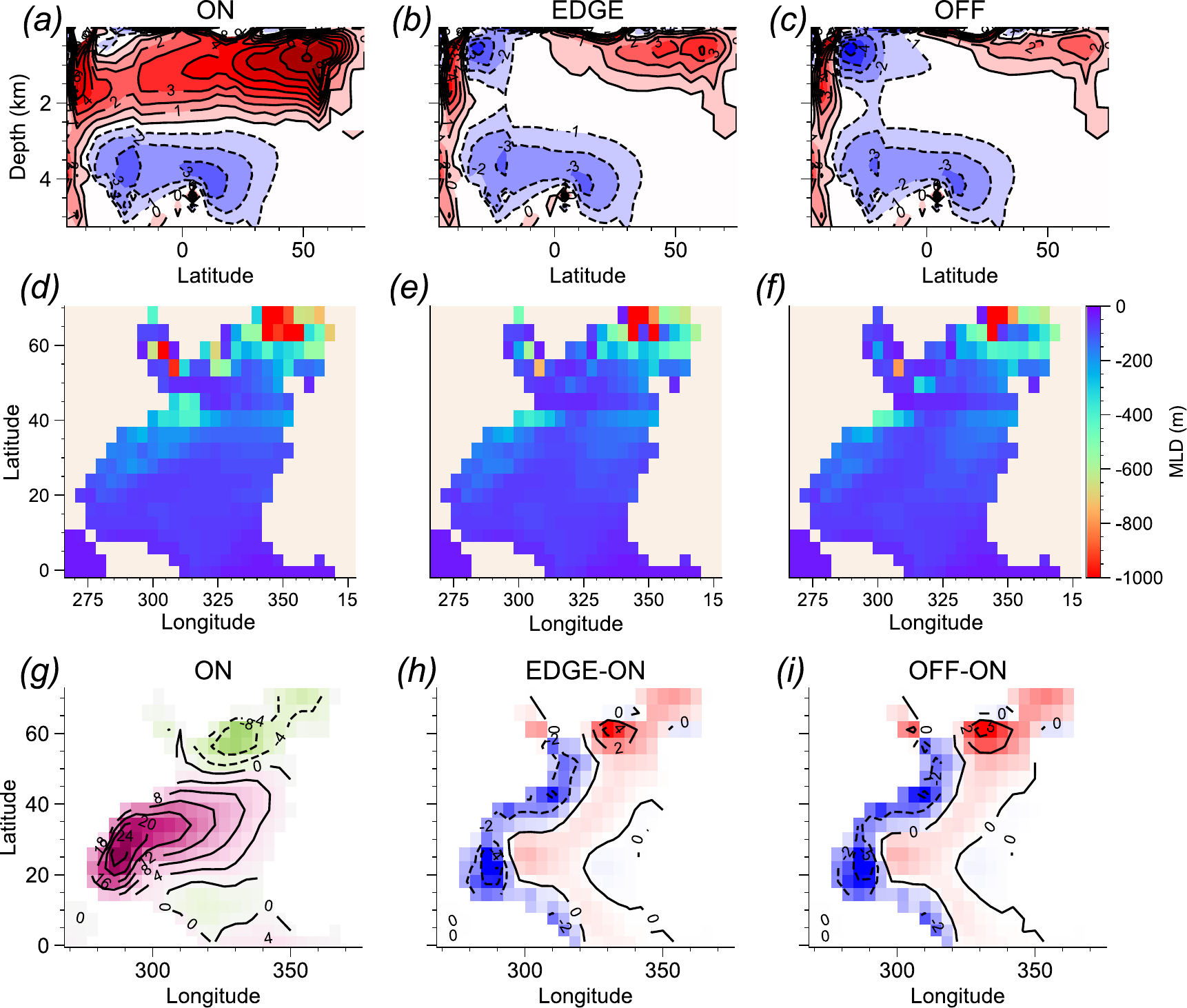}
\caption{\label{fig:circulation} 
{\it (a-c)} The Atlantic overturning streamfunction of the M state {\it (b)} is compared to those of the ON and OFF attractors. The contours show the circulation strength in Sv.
{\it (d-f)} Same, but for maps of the average winter mixed layer depth (MLD) in the NA. MLD is defined by the depth until which the monthly average ocean temperature is within 0.5 K of the sea surface temperature, indicating the degree of vertical mixing via convection. The winter MLD is obtained by averaging over the first 4 months of the year.
{\it (g-i)} Same, but for the barotropic streamfunction in the NA.
}
\end{figure}

We now compare the specific M state that was computed, which we will denote as 'EDGE' hereafter, to the ON and OFF attractors in terms of climatological averages. 
Estimating the M state from initial conditions in a different position in phase space (line L1 instead of L3) yields virtually identical results (see Appendix~\ref{AppB}). 
We consider the final 10 iterations of the edge tracking procedure and average the model state over the first 5 years of the final pairs of simulations. For the attractors we use the mean state obtained during the final 500 years of the long steady-state simulations. In terms of the NA circulation, EDGE lies 'between' ON and OFF featuring an intermediate strength of the AMOC and NA deep water formation (Fig.~\ref{fig:circulation}a-f). EDGE seems relatively closer to OFF compared to ON, for instance showing a weakening of both NA gyres that is almost as pronounced as the weakening in OFF (Fig.~\ref{fig:circulation}g-i, note positive anomalies of the barotropic streamfunction due to its negative sign in the subpolar gyre). 

\begin{figure}
\includegraphics[width=0.65\textwidth]{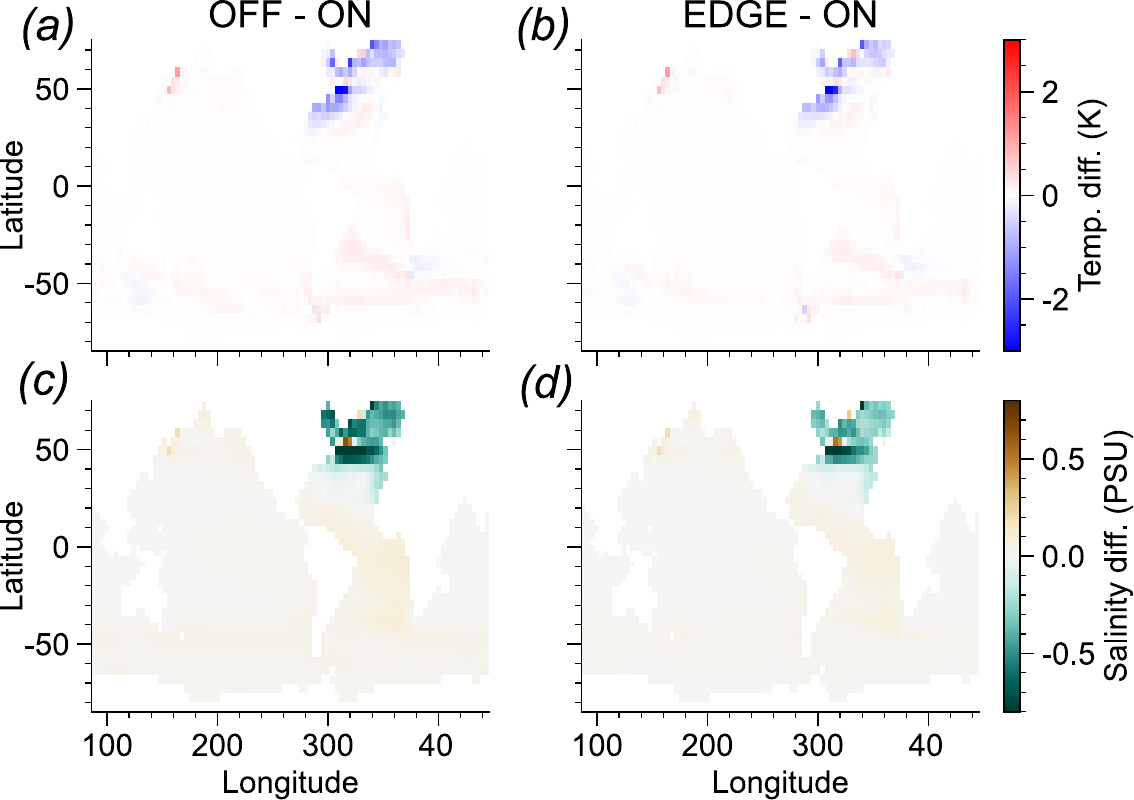}
\caption{\label{fig:surface_ocean} 
Anomaly in the sea surface temperature when moving from ON to OFF {\it (a)}, and 
from ON to EDGE {\it (b)}. {\it (c,d)} Same, but for sea surface salinity.
}
\end{figure}

Further, as opposed to ON, both EDGE and OFF lack any substantial inter-hemispheric transport (Fig.~\ref{fig:circulation}a-c). As a result, EDGE also shows strong cooling and freshening of the NA surface waters (Fig.~\ref{fig:surface_ocean}). Especially the temperature decrease is almost as pronounced as in OFF. This is not surprising since the AMOC is indeed severely weakened in EDGE, but it nevertheless confirms that the AMOC slowdown fingerprints - see  \citep{Roberts2013, Rahmstorf2015, Caesar2018, Jackson2020} - would also be observed in the case of an AMOC collapse via EDGE. 

\begin{figure}
\includegraphics[width=0.99\textwidth]{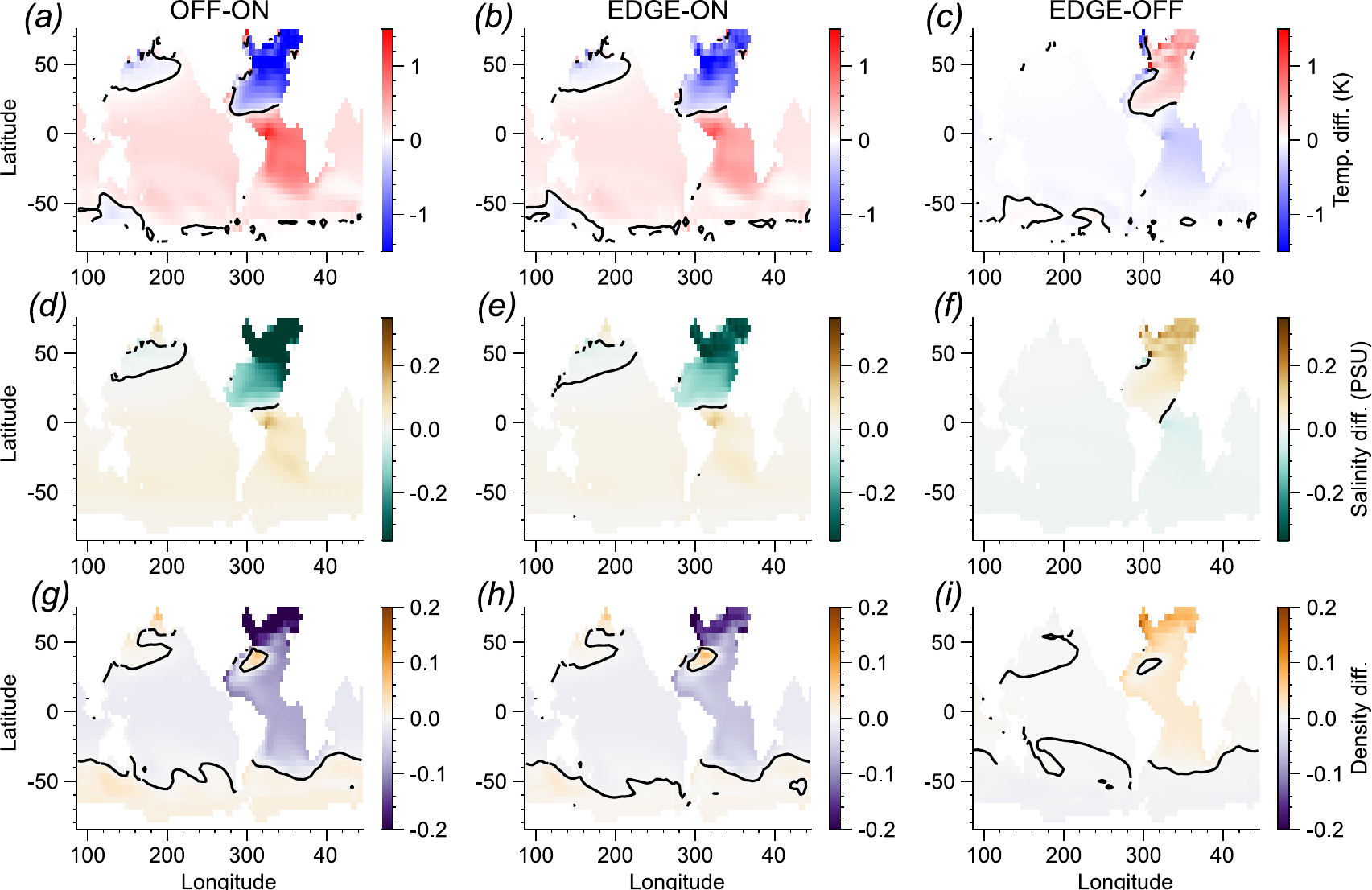}
\caption{\label{fig:global_upper_ocean} 
Differences between the attractors and the M state in the temperature {\it (a-c)}, salinity {\it (d-f)}, and density {\it (g-i)} averaged over the upper 1000~m of the ocean. 
}
\end{figure}

In the upper 1000 m of the global ocean the spatial pattern of the average temperature, salinity, and density anomalies of EDGE compared to ON are very similar to, but slightly weaker than those of OFF compared to ON (Fig.~\ref{fig:global_upper_ocean}a-d). Most significant is a NA cooling and freshening, and a SA warming and increase in salinity of EDGE and OFF compared to ON. Thus, in terms of the upper ocean properties, EDGE lies between ON and OFF. Going from ON to either OFF or EDGE shows clear a decrease in upper ocean density. Thus, as a consequence of an AMOC slowdown the NA freshening dominates the NA cooling, which supports that the destabilizing salt advection feedback is stronger than the stabilizing temperature advection feedback.

The situation in the deep ocean is markedly different, since here the properties of EDGE are outside of the range of the ON and OFF regimes for a large part of the global ocean. In particular, EDGE is colder and fresher than both ON and OFF in the entire Atlantic expect the Nordic seas, as seen by the same sign of the anomalies in Fig.~\ref{fig:global_deep_ocean}b,e and c,f. Due to the opposite effect of temperature and salinity, the pattern for the deep Atlantic density is more fragmented. In parts of the Atlantic freshening dominates over cooling, so that EDGE is less dense than both ON and OFF (Fig.~\ref{fig:global_deep_ocean}g-i). However, in other parts of the global ocean the density of EDGE lies between ON and OFF. 

\subsection{Fingerprinting the AMOC collapse}
\label{sec:fingerprint}

Having found a first indication of a characteristic anomaly in EDGE, we now want to consider how one can use the knowledge of such an anomaly to identify an ongoing AMOC collapse. Let us assume we are in the ON state and the AMOC is slowing down. We further assume that from climate modeling and paleoclimatic observations we have knowledge on the existence and features of an AMOC OFF state. We do not know whether the observed AMOC slowdown is only a temporary excursion, or whether it will lead us dangerously close to tipping. This needs to be assessed by consulting specific observables or fingerprints. Without specific knowledge of the exact transition path of an AMOC collapse - since it has not been directly observed before - it would be a priori assumed that the fingerprints pointing towards a possible transition from ON to OFF are in the direction of the ON-OFF anomaly, and that an ongoing transition should occur in the same direction. 
But from the theory of N-tipping and B-tipping 
we know that the relevant direction is instead determined by the position of EDGE with respect to ON. Hence, better fingerprints are those that are aligned in the ON-EDGE direction. While for some observables the ON-EDGE and ON-OFF anomalies point in the same direction, it is not the case for others. It is most important to identify the fingerprints where ON-EDGE and ON-OFF are not aligned, since in these cases the exclusive knowledge of the OFF state would lead one to believe that there is no dangerous excursion - or even an overall AMOC strengthening - because the system is moving further away from OFF. Note that if we are in the OFF state, the relevant direction for a transition is given by the EDGE-OFF anomaly. Hence, good fingerprints for the ON-to-OFF and OFF-to-ON transition are in general different. 

\begin{figure}
\includegraphics[width=0.99\textwidth]{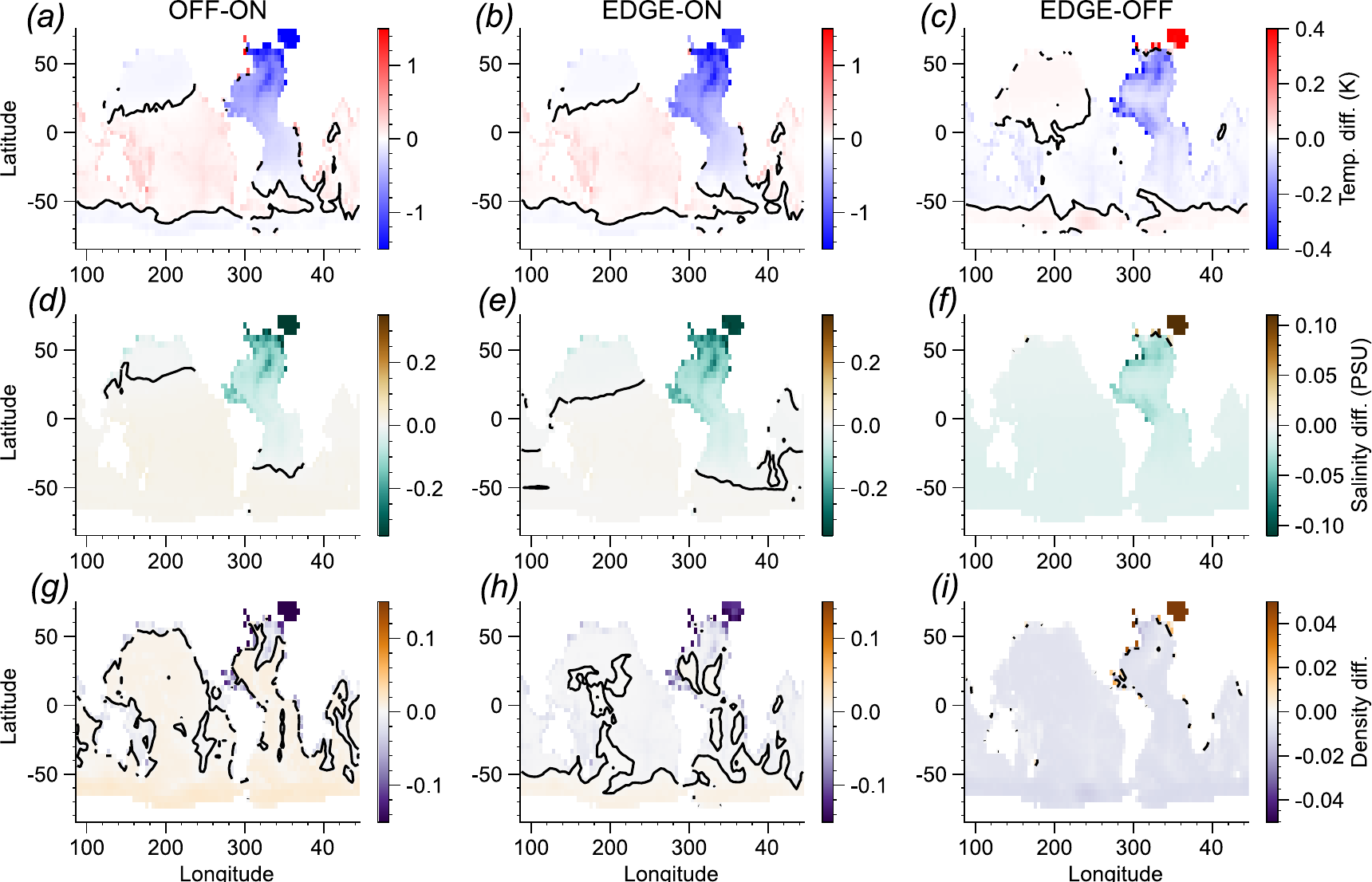}
\caption{\label{fig:global_deep_ocean} 
Same as Fig.~\ref{fig:global_upper_ocean}, but averaged for the ocean below 1000~m depth.
}
\end{figure}

With this in mind, for the structure of the deep Atlantic in EDGE we find that the ON-EDGE and ON-OFF anomalies point in the same direction, since OFF lies between ON and EDGE. Thus, there is a trend towards fresher and colder deep Atlantic both when expecting a hypothetical 'direct' transition from ON to OFF (Fig.~\ref{fig:global_deep_ocean}a,d), and when observing an excursion from ON towards EDGE (Fig.~\ref{fig:global_deep_ocean}b,e). The converse holds when considering an AMOC resurgence starting from OFF, because the OFF$\rightarrow$ON and OFF$\rightarrow$EDGE trends differ in sign. The AMOC resurgence scenario is not relevant for present-day climate, but its fingerprints may be useful for understanding abrupt changes in paleoclimate, such as the Dansgaard-Oeschger events. Outside the Atlantic, there are extended regions where the deep ocean density in ON lies between the densities in EDGE and OFF. Thus, in contrast to the Atlantic, in these regions the ON-OFF and ON-EDGE density anomalies point in different directions. Most notably, the density of the deep Indian ocean and almost the entire Pacific is lower in ON compared to OFF, but even lower in EDGE (Fig.~\ref{fig:global_deep_ocean}g,h). Here, a transition from ON via EDGE towards OFF would first see a decrease in density and then an increase. The initial density decrease is a non-trivial fingerprint of an AMOC collapse as it would not be expected from the knowledge of the OFF state only.

\begin{figure}
\includegraphics[width=0.8\textwidth]{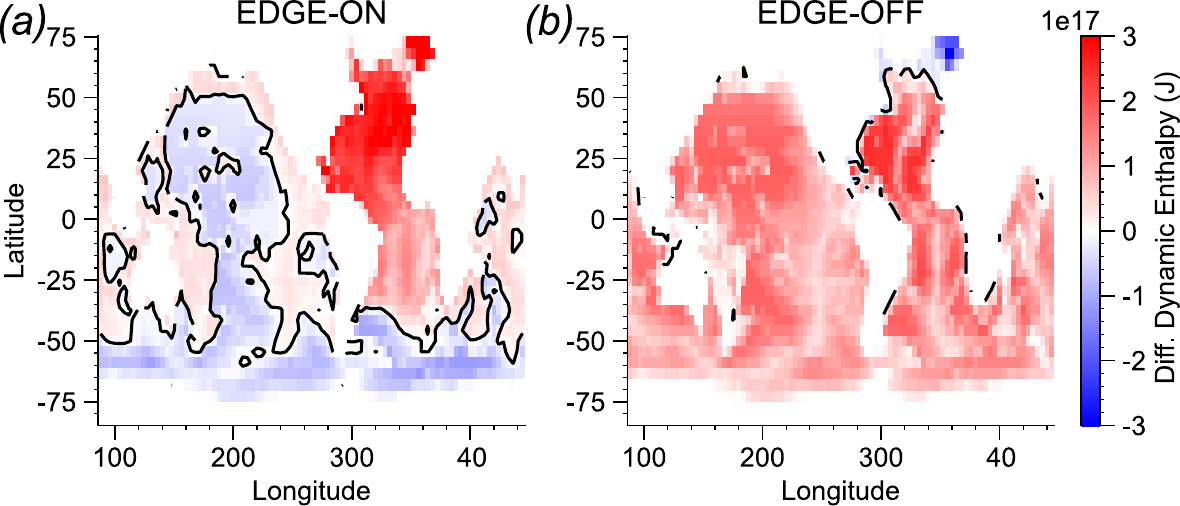}
\caption{\label{fig:enthalpy} 
Change in dynamic enthalpy (averaged over the entire water column) when moving from ON to EDGE {\it (a)}, and from OFF to EDGE {\it (b)}. In both cases, the global average is positive (see main text for numbers).
}
\end{figure}

When averaged over the ocean basins (disregarding the Southern ocean), the combined effect of temperature and salinity anomalies on the density is such that the deep ocean in EDGE is lighter compared to both ON and OFF (Fig.~\ref{fig:global_deep_ocean}h,i). The lighter waters at depth raise the center of gravity, and thus EDGE has the highest gravitational potential energy of the three states. This can be quantified by computing the dynamic enthalpy \cite{YOU10}, which is the effective potential energy of the ocean and measures the energy released or required 
when moving a fluid parcel of unit mass adiabatically from its current position to the surface \cite{NYC10}. 
The dynamic enthalpy for a parcel at depth $z$ is referenced to the surface, yielding a negative value. Its variations depend on the mean and vertical distribution of density of the ocean column above. 
Compared to the attractors, in EDGE there is on average less energy required to move parcels up and thus it represents a higher potential energy state. The vertically averaged dynamic enthalpy shows that for the entire ocean except the Nordic seas EDGE is locally in a higher energy state compared to OFF (Fig.~\ref{fig:enthalpy}b). This corresponds well with the consistently lower deep ocean density in EDGE (Fig.~\ref{fig:global_deep_ocean}i), and it yields a global dynamic enthalpy that is 2.09$e^{20}$J higher in EDGE compared to OFF. 
When comparing EDGE to ON, the deep ocean density anomaly is more fragmented (Fig.~\ref{fig:global_deep_ocean}h), which leads to a depth-averaged dynamic enthalpy that is consistently higher in the Atlantic only (Fig.~\ref{fig:enthalpy}a). But due to the strong Atlantic anomaly the globally averaged dynamic enthalpy in EDGE is still higher by 4.86$e^{19}$J. 

\begin{figure}
\includegraphics[width=0.95\textwidth]{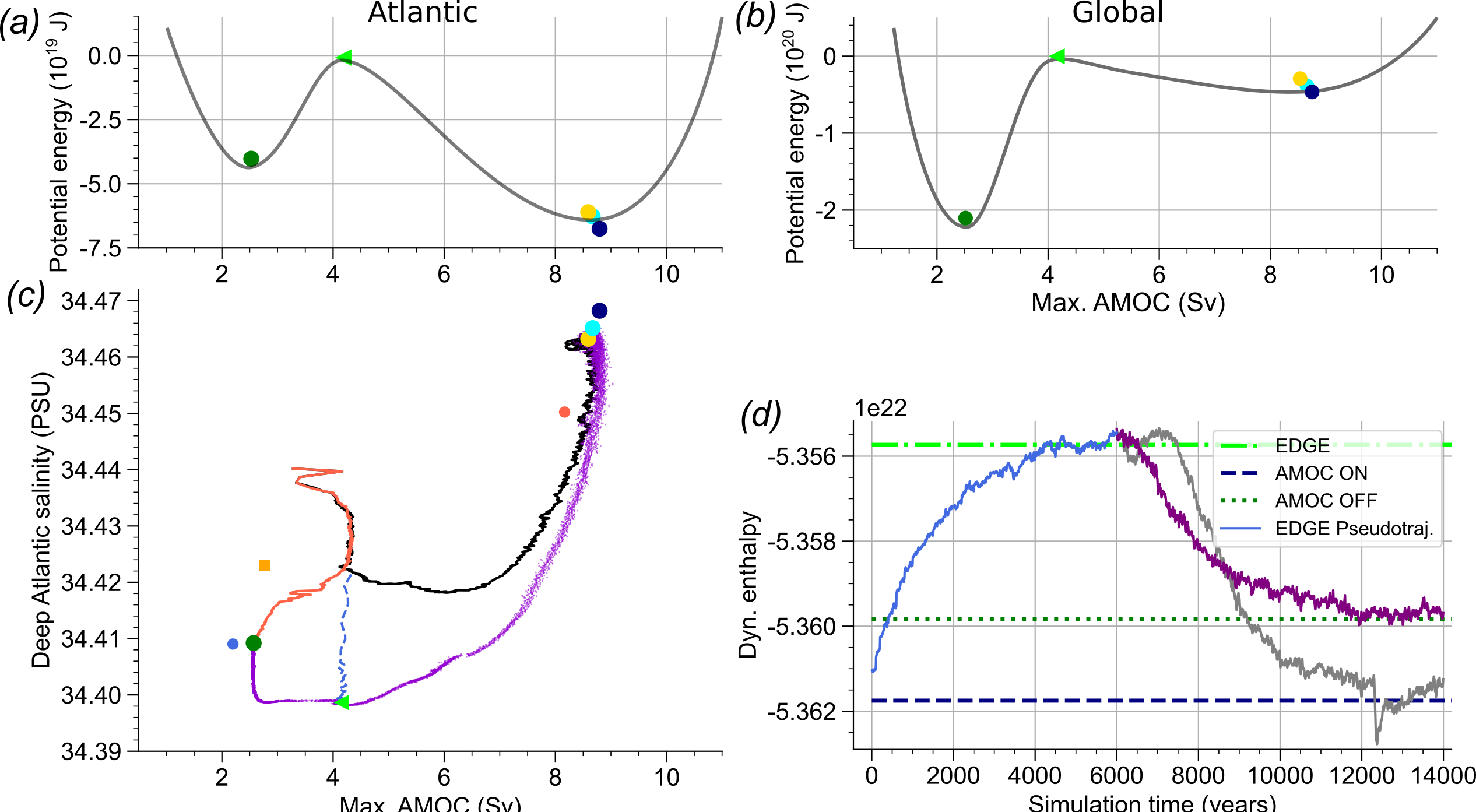}
\caption{\label{fig:summary_fig} 
{\it (a,b)} Conceptual illustration of quasipotential landscapes projected onto the AMOC strength, derived from the dynamic enthalpy integrated over the {\it (a)} Atlantic and {\it (b)} global ocean in the attractors and the M state. The values have been offset such that the potential energy in the M state is 0. 
{\it (c)} Two-dimensional projection of the mean states in the attractors (colored dots) and the M state (green triangle), as well as the basin boundary (blue dashed line) and the relaxation trajectories (purple), computed by extending the simulation time for the pairs of trajectories during the 10 last iterations of the edge tracking algorithm to 4~000 years. 
{\it (d)} Time average of the total Atlantic dynamic enthalpy in the attractors and M state (horizontal lines). The blue time series is a pseudo-trajectory approaching the M state during the edge tracking algorithm, obtained by splicing together the 100-year paths (given by $t_1$) that correspond to the movement along the basin boundary for subsequent iterations. Two trajectories that relax from the vicinity of the M state (starting at $t=6000$~years) to the ON and OFF attractors are also shown. 
}
\end{figure}

Based on this, we draw an analogy between the quasipotential of the underlying dynamical system and the (physical) potential energy of the ocean. The higher dynamic enthalpy of the unstable M state compared to the attractors corresponds well with the latter as local minima in the quasipotential landscape. 
We illustrate this in Fig.~\ref{fig:summary_fig}a-c, where a two-dimensional phase space projection is given together with conceptual quasipotential landscapes represented by the dynamic enthalpy. The common feature concerning both transitions starting at ON or OFF  is that potential energy needs to be gained in the Atlantic in order to reach the M state (Fig.~\ref{fig:enthalpy}). Thus, in Fig.~\ref{fig:summary_fig}a we give a simplified representation of the potential barrier that needs to be overcome in terms of the dynamic enthalpy in the Atlantic (north of 30$^o$S). For a transition in either direction, this is mainly achieved by bringing relatively fresh water to the deep Atlantic (Fig.~\ref{fig:global_deep_ocean}e,f). Starting at the attractors, this corresponds to a movement in the negative vertical direction in Fig.~\ref{fig:summary_fig}c, where we also show relaxation trajectories (purple) from the vicinity of EDGE towards the attractors. These are used as stand-in for the instantonic trajectories of a typical noise-induced transition, but these are generally not the same in nonequilibrium systems \citep{Graham1987,Graham1991,Lucarini2020,Margazoglou2021}. For a transition from OFF, potential energy needs to be gained not only in the Atlantic but also everywhere else (again by deep ocean freshening, see Fig.~\ref{fig:global_deep_ocean}f). As a result, the magnitude of the total potential difference is better represented by a conceptual quasipotential landscape based on the {\it globally} integrated dynamic enthalpy (Fig.~\ref{fig:summary_fig}b). 

Additional details of the transition mechanism to the M state may be missing if the instantons, and specifically their initial segments beginning at the attractors, may point yet in another direction in phase space compared to anomaly of the M state. Computing instantonic trajectories is beyond the scope of our study, because it requires either running lengthy stochastically forced simulations and then performing suitable stochastic averages \citep{Margazoglou2021}, or solving a complex variational problem \citep{Grafke2019}. 
After reaching the M state, the instanton is expected to follow the relaxation trajectory towards the respective attractor. Here, potential energy is released again by transfer of salty water to the deep Atlantic (positive vertical direction in Fig.~\ref{fig:summary_fig}c). 


We used the Atlantic dynamic enthalpy to illustrate the physics of EDGE as a state of higher potential energy compared to both ON and OFF. Note, however, that since the Atlantic domain is an open system, its dynamic enthalpy is by no means a true Lyapunov function of the (non-equilibrium) system that would be strictly minimized in a stable steady state. 
Trajectories starting at EDGE show the expected roughly monotonic decrease in the potential, as seen by two trajectories (purple and gray time series in Fig.~\ref{fig:summary_fig}d) leading to the competing attractors. But we also find initial conditions where the deterministic evolution leads to temporary, but significant increases in potential energy, while we should only see a monotonic decrease along the free deterministic flow. A more serious flaw is that EDGE does not appear to be a saddle for the chosen quasipotential. For a saddle the potential should decrease as we approach EDGE along its stable manifold on the basin boundary, and further decrease when leaving along its unstable manifold. However, in our case the dynamic enthalpy increases monotonically during the movement along the basin boundary, as can be seen by the pseudo-trajectory along the basin boundary during edge tracking (blue time series Fig.~\ref{fig:summary_fig}d). Hence, EDGE is instead a local maximum of the dynamical enthalpy. 



\section{Discussion}
\label{sec:conclusions}


In this work we have investigated the global stability properties of the AMOC by focusing, rather than on the competing attractive circulation regimes, on the properties of the dynamical saddles separating such states.

We harness the edge tracking algorithm to compute the M state, an unstable AMOC configuration that lives on a basin boundary separating two attracting regimes of a vigorous and collapsed circulation in a global ocean model. The main physical relevance of such an unstable state is its potential role as gateway of a noise-induced collapse of the AMOC. Determining its physical characteristics uncovers the oceanographic changes to be expected leading up to a noise-induced collapse. Since a priori the M state could be anywhere in phase space, the AMOC strength could, for instance, first increase during an ongoing transition to the OFF state, making an early detection non-obvious. This specific scenario seems less likely with our results. Information on the M state climatology can also be used to identify the physical mechanisms that are necessary to drive a collapse, as well as specific perturbations that would greatly increase the risk of a transition into the competing basin of attraction. Furthermore, when the forcing parameter is increased towards the tipping point where the reference attractor loses stability, fluctuations away from the attractor are increasingly directed towards the M state (along the instanton). This may inform us about observables, the increases in fluctuations of which could serve as early-warning signals.

By analyzing its climatology, we find that the M state lies consistently between the ON and OFF states in terms of the upper ocean properties. Here it features a freshening and cooling of the NA (and vice versa for the South Atlantic), which is a known fingerprint of an AMOC slowdown. But the deep ocean properties in EDGE are distinct and outside the range characterizing ON and OFF. Most notably, EDGE features a cooling and freshening of almost the entire Atlantic compared to both ON and OFF regimes. In oceanographic terms, EDGE is less 'spicy' (warm and salty) in the deep Atlantic compared to the attractors. Since the OFF state lies between ON (most spicy) and EDGE (least spicy), a transition from ON to EDGE and from ON to OFF both feature a decreasing trend in spiciness. In contrast, in the deep ocean outside the Atlantic there is a decrease in density when moving from ON to EDGE, which is opposite to the trend when comparing ON and OFF. Thus, in terms of this observable, a dangerous transition towards EDGE is the opposite as would be expected when only knowing the ON-OFF anomaly and using the latter to construct observables that signify an expected AMOC collapse. This shows the importance of identifying all relevant characteristics of the M state (in addition to OFF) in order to avoid a misinterpretation of oceanographic changes leading up to an AMOC collapse. This is especially relevant as direct observations of the AMOC strength have only been performed for the last two decades, which is too short to detect changes considering the large variability. 

It may, however, be questioned whether deep ocean fingerprints are useful in practice, due to difficulties in obtaining observations, and more fundamentally because processes on very long time scales are likely involved to accomplish the substantial change of the deep ocean properties (freshening and cooling of the deep Atlantic beyond the values in OFF) required to reach EDGE. It may thus be unlikely that a spontaneous collapse of the AMOC would indeed occur as a passage via the instanton in the low-noise limit. Instead, it may be a more likely scenario that, on shorter time scales, finite-amplitude noise can kick the AMOC over the basin boundary in places within a relatively short distance of the latter but far away from the M state. This could be especially true if the noise has a considerable jump process component \citep{Lucarini2022}. Due to the specific deep ocean fingerprint, the likelihood of a transition via EDGE may also be quite different for an AMOC collapse compared to an AMOC resurgence. If the AMOC collapse is initiated by the cessation of NA deep water formation, as commonly assumed, the Atlantic deep ocean properties would not immediately change towards the EDGE conditions. This would need to happen gradually over time, possibly by processes outside the NA. The AMOC resurgence, on the other hand, may occur by a spontaneous onset of NA deep convection, which could in principle flush the deep ocean with less spicy water much quicker and change its properties towards EDGE. 

The characteristics of the AMOC M state presented here need to be tested in models with higher spatial resolution. Our coarse-resolution ocean model does not resolve turbulent eddy dynamics, which may have an impact on the mean ocean state and its stability. 
For instance, mesoscale eddies significantly influence the strength and variability of deep water formation \citep{GEL11,KAT18,BRU19}, which in turn is a strong control on the AMOC. 
How the presence of eddies impacts the anatomy of the AMOC TP and the M state that lies between the ON and OFF regimes needs to be tested in a higher resolution setup. This would require significantly higher computational resources, both in order to perform the edge tracking and to establish the stability landscape as an initial step. 
After establishment of the basins of attraction and the boundary, the computation of the edge state alone, based on two initial conditions and using $M=60$ iterations with $k=7$ bisections and a simulation time of $t_3=500$~years, required 420,000 model years, which can only be minimally parallelized because a given realization needs to be simulated until $t_2 = 250$~years before starting the next realization, in order to decide whether it tips up or down. 

Further, our ocean-only model lacks processes on fast time scales due to atmospheric feedbacks and dynamical sea ice. The latter is only implemented in a simplified way by its insulating effect (see Appendix~\ref{Appe}), which is, however, an important control on the hysteresis of the AMOC \citep{VWE24b}. The coupled climate system could in principle feature qualitatively different characteristics of the M state and AMOC transition pathways, which possibly evade the long time scales associated with the deep ocean. 
This needs to be tested with a coupled atmosphere-ocean model, which is computationally expensive. It also remains to be understood better whether the coupled climate system in the context of the AMOC can indeed be considered as a system where the deterministic ocean is forced by an effectively stochastic atmosphere that operates in the low-noise limit such that the large-deviation theory applies, and as a result, where a transition passes via the M state. 

The Graham field theory indicates that an M state is a saddle in the quasipotential landscape. While we have not derived a quantity that can serve as quasipotential or Lyapunov function in the strict sense, EDGE indeed features higher available potential energy compared to the attractors, as measured by the relative magnitudes of the dynamic enthalpy. In a transition between the attractors via EDGE, potential energy needs to be gained, which, in the case of the M state we identified, needs to be achieved by bringing relatively fresh (and thus light) water to the deep Atlantic. It would be interesting to compare the measure of potential energy employed here to empirical quasipotentials obtained under stochastic forcing \citep{Lucarini2019,Lucarini2020,Margazoglou2021}. By performing experiments at different values of the control parameter, it could then be investigated how changes in EDGE, and specifically the potential energy difference of attractors and EDGE, may inform us of the likelihoods of spontaneous transitions, as suggested by Arrhenius' law. Changes in the edge state may also concern its dynamical features, as opposed to the mean state discussed here, which could lead to more useful fingerprints of an approaching AMOC collapse.


An important fundamental aspect that we have been unable to address here is the study of the multitude of M states that may exist as our system features several (at least seven) competing states for the chosen value of the control parameter $F$. We have been able to only very partially explore the complexity of the basin boundaries, each potentially supporting one or more M states, which could in principle lead to a variety of distinct pathways for AMOC tipping. The possibility of so-called Wada basins \citep{KEN91}, where more than two basins of attraction share one boundary, makes an investigation into the unknown number of co-existing M states even more intriguing. We have so far not been able to find other M states that connect the vigorous and collapsed AMOC regimes, since the only basin boundary we could locate was between the two specific attractors 'ON' and 'OFF'. But additional M states may nevertheless exist and this should be investigated with more extensive ensemble simulations or other techniques.

\textbf{Acknowledgements} We thank R. Nuterman and the Danish Center for Climate Computing for supporting the simulations with the Veros ocean model.

\textbf{Funding} JL has received support from Danmarks Frie Forskningsfond under grant no. 2032-00346B. VL has received support from the EU Horizon Europe project ClimTIP (Grant No. 101137601), from the Marie Curie Action CRITICALEARTH (Grant No. 956170), and from the EPSRC project LINK (Grant No. EP/Y026675/1).

\textbf{Declaration of interests} The authors report no conflict of interest.

\textbf{Data availability statement} The raw model simulation data that underlie
the findings of this study are deposited in the Electronic Research Data Archive repository of the University of Copenhagen and can be accessed at https://sid.erda.dk/sharelink/BMov9nurth.

\textbf{Author ORCIDs} J. Lohmann: https://orcid.org/0000-0002-6323-6243; V. Lucarini: https://orcid.org/0000-0001-9392-1471.

\textbf{Author contributions} JL and VL conceived the study and designed the numerical experiments. JL performed the numerical simulations and data analysis. JL and VL wrote the paper.

\bibliography{refs} 

\appendix 
\renewcommand\thefigure{\thesection.\arabic{figure}}    

\section{Details on the ocean model}\label{Appe}

We employ the primitive equation finite-difference global ocean model {\it Veros}, a direct translation of the Fortran backend of the ocean model {\it PyOM2} \citep{EDE14} into Python/JAX \citep{HAE18,HAE21}. Mesoscale turbulence is represented using the \citet{RED82} and \citet{GEN90} parameterization for isopycnal and thickness diffusion (diffusivity of 1~000 m$^2$/s). We use the second order turbulence closure of \citet{GAS90} to account for diapycnal mixing. Here, a background diffusivity of 10$^{-5}$ m$^2$/s is employed, and static instabilities are removed by a gradual increase of the vertical diffusivity as the instability is approached. This corresponds to a smoothed variant of convective adjustment. 

The heat exchange boundary condition is expressed by a first-order Taylor expansion of the heat flux as a function of the anomaly of the modeled surface temperature with respect to a fixed surface temperature climatology \citep{BAR98}. For this, we use ERA-40 climatologies \citep{UPP05} of surface temperature and heat flux, as well as a climatology of the derivative of the heat flux with respect to changes in surface ocean temperature. The latter is derived from ERA-40 following \citet{BAR95} and has a global yearly average 30.26 $WK^{-1} m^{-2}$. Wind stress forcing is also taken from the ERA-40 reanalysis data. Freshwater exchanges with the atmosphere are modeled by boundary conditions under which the sea surface salinity is relaxed towards a present-day ERA-40 climatological field within a given relaxation timescale.  By choosing a long timescale of 2 years, oceanic salinity anomalies are less efficiently damped by the atmospheric forcing compared to temperature anomalies. This enables the positive salt advection feedback, which can lead to AMOC multi-stability and tipping. 

The representation of sea ice in Veros highly simplified, as the ocean is not coupled to a dynamical sea ice component. However, the important insulating effect of advancing sea ice is incorporated. When below the typical freezing point of ocean water (-1.8~deg~C), the heat and freshwater exchange with the atmosphere at the given surface grid cell is disabled. 

The bathymetry is obtained by smoothing the ETOPO1 global relief model \citep{AMA09} with a Gaussian filter to match the grid resolution. A current limitation of the model is its simple rectilinear grid, which implies that the Arctic pole cannot be modeled due to the vanishing grid size. Thus, the model domain ends at 80$^o$N  with no Arctic connection of the ocean basins. 
The horizontal grid contains 90 longitudinal and 40 latitudinal cells, where the latitudinal resolution increases from 5.3$^o$ at the poles to 2.1$^o$ at the Equator, as well as 40 vertical layers, which increase in thickness from 23 m at the surface to 274 m at the bottom. For further model details, see our previous studies that investigated AMOC tipping in this ocean model \citep{LOH21,LOH22,Lohmann2024}.

In addition to the ERA-40 forcings, a freshwater anomaly is introduced in the North Atlantic, in order to drive the system closer to the tipping point of a collapsed AMOC.
To this end, a salinity flux anomaly $\tilde{\phi}$ is applied at the surface in the grid cells between 296$^o$W to 0$^o$W and 50$^o$N to 75$^o$N, corresponding to an area of about $A= 1.5 \, \text{mio.} \, km^2$. Here we use the equivalent total freshwater forcing $F = \tilde{\phi} A S_{ref}^{-1}$ as control parameter, with the reference salinity $S_{ref} = 3.5 g\cdot kg^{-1}$.

\section{M state estimated from different initial conditions}\label{AppB}
\setcounter{figure}{0}    

\begin{figure}
\includegraphics[width=0.99\textwidth]{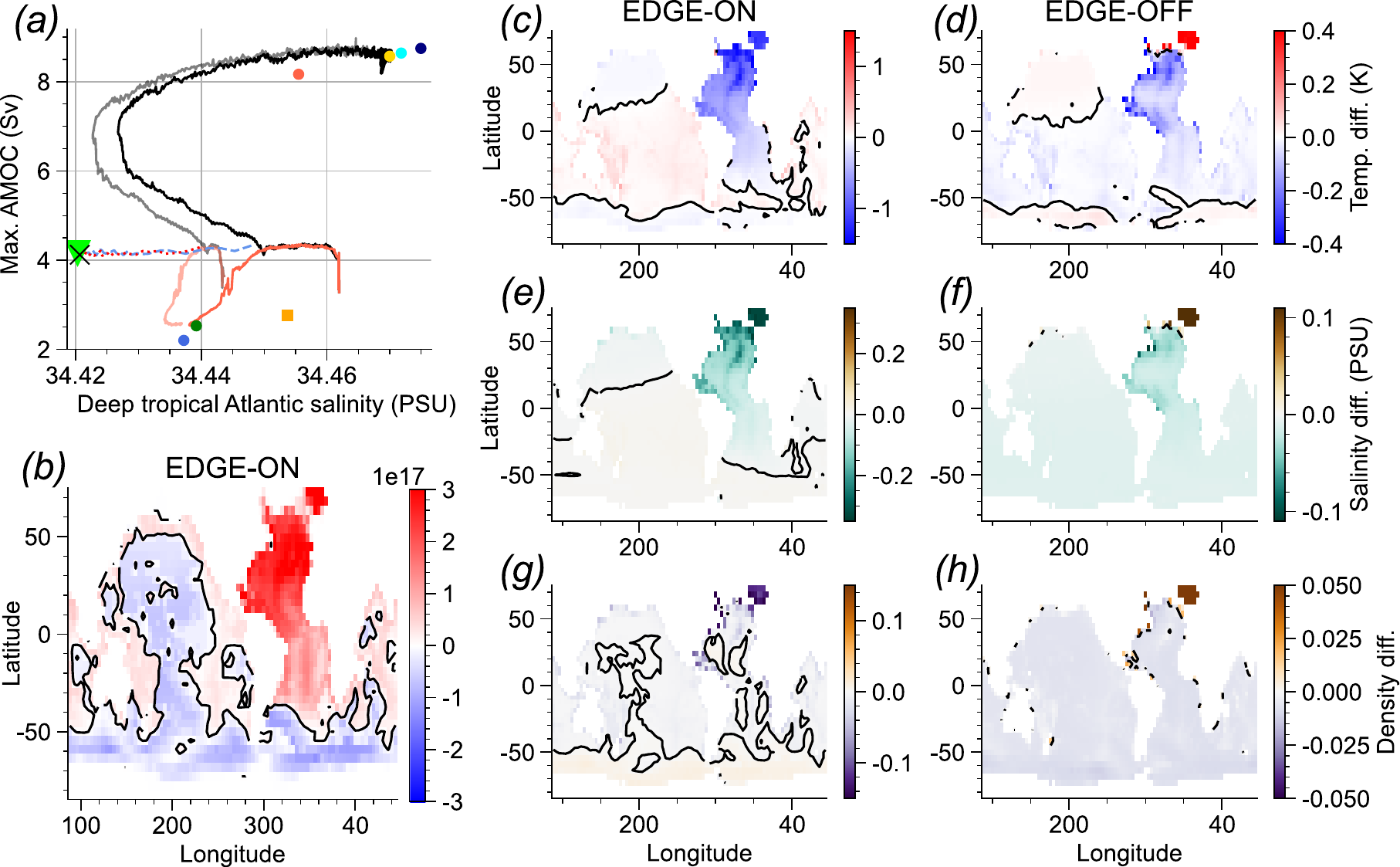}
\caption{\label{fig:edge_alternativ} 
{\it (a)} Projection of the M states (green triangle and black cross) and basin boundaries (red dotted and blue dashed lines) estimated by the edge tracking procedure for two different initial conditions. The respective initial conditions are taken by bisecting two very nearby trajectories (shown in red and black) on opposite sides of the basin boundary shortly after they separate. The two trajectories for the alternative initial conditions are shown with reduced contrast. Also shown are the positions of the co-existing attractors (colored symbols). 
{\it (b)} Anomaly of the depth-integrated dynamic enthalpy of the M state with respect to the ON state. 
{\it (c-h)} Depth-integrated temperature, salinity and density anomalies in the ocean below 1000 m of the M state with respect to the ON and OFF attractors.
}
\end{figure}

The M state we described in Sec.~\ref{sec:edge_tracking}-\ref{sec:fingerprint} appears to be a robust steady state of the system. Within statistical uncertainties, the same state is reached when starting the edge tracking algorithm from two very close-by states on an entirely different location of the basin boundary. In order to provide support to this, we chose the intersection of the boundary and the straight line L1 (as opposed to L3), which is defined by a pair of different attractors compared to the results presented above. Selected properties of the model state obtained after 50 iterations of edge tracking using these initial conditions are shown in Fig.~\ref{fig:edge_alternativ}. The shadowed states on the basin boundary (red dotted line), as well as the estimated M state (black cross) match the previous results very well (Fig.~\ref{fig:edge_alternativ}a), and the characteristic deep ocean anomaly patterns are near-identical (compare Fig.~\ref{fig:edge_alternativ}b-h to Fig.~\ref{fig:enthalpy}a and Fig.~\ref{fig:global_deep_ocean}). 

\end{document}